# Predicting Defect Content and Quality Assurance Effectiveness by Combining Expert Judgment and Defect Data - A Case Study

Michael Kläs[1], Haruka Nakao[2], Frank Elberzhager[1], Jürgen Münch[1]
[1]*Fraunhofer Institute for Experimental Software Engineering, Germany*
[2]*Safety & Product Assurance Dept. Japan Manned Space Systems Corporation, Japan*
michael.klaes@iese.fraunhofer.de, haruka@jamss.co.jp, {frank.elberzhager, juegen.muench}@iese.fraunhofer.de

**Abstract**

*Planning quality assurance (QA) activities in a systematic way and controlling their execution are challenging tasks for companies that develop software or software-intensive systems. Both require estimation capabilities regarding the effectiveness of the applied QA techniques and the defect content of the checked artifacts. Existing approaches for these purposes need extensive measurement data from historical projects. Due to the fact that many companies do not collect enough data for applying these approaches (especially for the early project lifecycle), they typically base their QA planning and controlling solely on expert opinion. This article presents a hybrid method that combines commonly available measurement data and context-specific expert knowledge. To evaluate the method's applicability and usefulness, we conducted a case study in the context of independent verification and validation activities for critical software in the space domain. A hybrid defect content and effectiveness model was developed for the software requirements analysis phase and evaluated with available legacy data. One major result is that the hybrid model provides improved estimation accuracy when compared to applicable models based solely on data. The mean magnitude of relative error (MMRE) determined by cross-validation is 29.6% compared to 76.5% obtained by the most accurate data-based model.*

## 1. Introduction

Quality assurance (QA) is an essential part of today's software projects. This holds in particular when dependable software-intensive systems are being developed, where delivering high quality is a major success factor. The reduction of quality risk achieved by performing QA activities usually consumes a large portion of the project budget (between 30 and 90 percent) [1]. Therefore, the accurate planning and controlling of QA activities (in particular QA activities during early software lifecycle phases like requirements reviews and inspections) can contribute significantly to the project's success in terms of quality and project cost.

When planning QA activities in a project, one major question is whether the planned QA activities are appropriate for reducing the product-specific quality risk to an acceptable level, i.e., whether the planned activities are effective enough to handle the expected defect content of the checked product. The next issue is to determine whether the planned level of risk reduction has been achieved. Thus, indicators or thresholds for the expected number of defects found by the planned QA activities are required.

However, predicting the effectiveness of a planned QA activity or the defect content of an artifact is not a trivial task. Earlier research focused on finding a single size or complexity measure as an accurate predictor of defect content [2], [3]. Another approach was to provide a fixed effectiveness value or range for certain QA techniques by conducting empirical or simulation studies. In practice, both led to disappointment: Single size and complexity metrics are recognized as being insufficient predictors [4] and the effectiveness of QA techniques varies extensively depending on the concrete context of the study [5], [6]. Therefore, effectiveness values described in the literature cannot be directly applied as predictors for QA effectiveness in its own context.

From our point of view, the reason for the observed problems is that the effectiveness of QA activities as well as the defect content of checked artifacts are complex constructs, which are (1) very context-specific and (2) influenced by several factors. A reasoned literature research [7] identified over 100 different factors mentioned across the software engineering literature that have an impact on defect content or QA effectiveness.

Newer research directions are tackling the problem by using multivariate approaches like [8] and [9], which allow building context-specific models and consider more than one independent variable (i.e., influencing factor). These approaches are applied successfully in areas where huge datasets are available (more than 100 data points) or where collection without major effort is possible, respectively. For example, [10] and [11] used automatically collected code metrics to successfully predict defect-prone modules.

However, practical application of these models is limited by their data requirements, which holds especially for early software development phases, where no automated data collection is available. For example, only few companies have access to collected measurement data about the effectiveness of inspections and relevant influencing factors for more than 100 requirements inspections (i.e., data points).

Consequently, we see a lack of empirically validated methods that can be used to build context-specific prediction models for defect content and effectiveness when limited quantitative data is available.

Considering the field of cost estimation, methods are described for building and validating context-specific prediction models, even in cases where few data points and limited measurement data are available. Methods like CoBRA [12] have been applied successfully in many case studies to build context-specific prediction models using expert experience and existing measurement data [12], [13], [14]. Therefore, we adopt the idea of combining expert opinion with measurement data in a rigorous manner and adapt it for defect content and QA effectiveness prediction. The method we present supports the planning of QA activities by providing assessments of the remaining quality risks related to the planned QA activities and providing thresholds for controlling the planned QA activities.

To evaluate the proposed method with respect to applicability, application cost, and usefulness, a case study [15] was performed. The study was conducted in the context of software Independent Verification and Validation (IV&V) performed for mission-critical space systems. Independence is considered here in terms of technical, managerial, and financial independence [16].

The paper is organized as follows: Section 2 discusses related work regarding planning and controlling early quality assurance techniques. Section 3 presents the HDCE method, which allows building a hybrid prediction model combining expert opinion and measurement data for planning and controlling QA activities. In Section 4, the method is evaluated in a case study with respect to applicability and application cost. A set of context-specific influencing factors is presented, and the usefulness of the resulting model is evaluated by applying the model on legacy data. Finally, Section 5 discusses threats to validity and Section 6 concludes and outlines directions for future research.

## 2. Related Work

While considering the planning and controlling of quality assurance techniques applied during early lifecycle phases, we identified various techniques. One approach to controlling inspections is the use of capture-recapture models [17]. They answer the question of whether inspections can be stopped or should be continued in terms of the number of defects remaining within a document. For this purpose, an estimation of the total number of defects within a document is performed based on the defects already found by different inspectors. A calculation of the effectiveness and the estimated number of remaining defects can be done based on the assumption that a large amount of defects found by only one inspector results in a high overall defect count. If many defects are found by more than one inspector, the overall number of defects is estimated to be low. The main problem is the number of inspectors required to obtain suitable estimation results, which is a minimum of 4.

Another approach to controlling inspections are curve-fitting techniques as described by Wohlin et al. [18], which are derived from ideas of reliability growth models used for controlling testing activities. Here, defect data is gathered for inspection activities and different distributions are calculated to predict the overall number of defects. The accuracy of such models was not as high as that of capture-recapture methods [19].

For the second application mentioned – planning inspections – only few models exist. Previous research mainly focused on controlling inspections. One approach to planning inspections is described by Briand [20]. Data gathered from about 150 inspections are used to perform a linear regression analysis, considering the size of the artifact and the preparation effort. The main problem with this approach is the extensive number of data points necessary for performing such an analysis. Finally, MARS (Multiple-Adaptive-Regressions-Splines) has shown its suitability for planning, compared to different regression analyses, based on three studies [21]. However, MARS also requires many data points (>100 inspections) for valuable application [21].

Consequently, the main problem regarding existing methods is, on the one hand, the need for a large number of inspectors for suitable predictions and, on the other hand, a huge amount of necessary data gathered for inspections, which is often not available.

## 3. The HDCE Method

The <u>H</u>ybrid <u>D</u>efect <u>C</u>ontent and <u>E</u>ffectiveness method (HDCE) combines expert judgment and available measurement data from current and historical projects to provide guidance for QA planning and controlling. The idea of combining expert opinion and measurement data supported by a quantitative causal model and Monte Carlo simulation is taken from the cost estimation area [12] and adapted to defect content and effectiveness prediction.

This section first describes the *HDCE model* as the core element of the HDCE method, then gives an overview of the process employed for building the *quantitative causal model* using expert judgment, and, finally, explains how the method can be applied to QA planning and controlling.

### 3.1 Hybrid Defect Content & Effectiveness Model

This section describes the hybrid defect content and effectiveness model. Figure 1 provides a general overview of the relationship between the components in the HDCE model: The *DCE equation* combines the *defect content* and the *effectiveness model*; the *quantitative causal model* captures the expert opinions; the *historical project data* are used to derive a defect content and effectiveness baseline; and the *characterization of the actual project* allows determining the project-specific *DDIF and EIF distributions* with the help of *Monte Carlo simulation*.

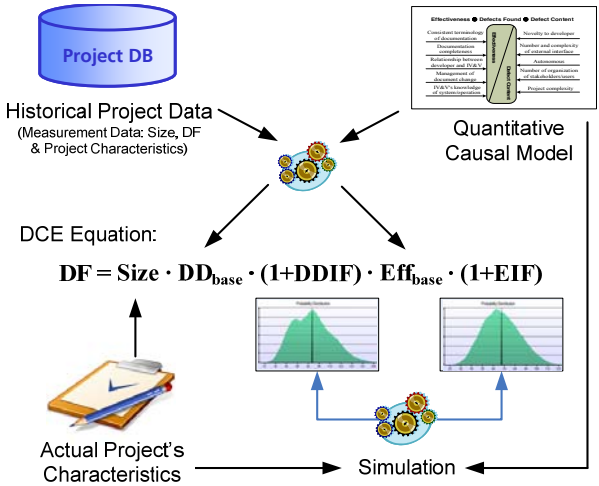

**Figure 1. Overview of HDCE model**

### 3.1.1. Defect Content and Effectiveness (DCE) Equation

The number of *defects found* (DF) by a QA activity can be described by two variables, the number of defects in the checked artifact at the moment when the QA activity was performed (i.e., the defect content) and the effectiveness of the QA activity itself (i.e., the ratio between the number of defects found and the total number of defects in the product when the QA activity is conducted [22]):

**Defects Found = Defect Content · Effectiveness   (4)**

In the following, we describe one model for defect content and one model for QA effectiveness. Equation (4) describes how these models are related and Figure 1 shows the resulting defect content and effectiveness equation as the glue between the other parts of the HDCE model.

*Defect Content Model:* In contexts where a well-established development process is used and the development team is homogenous and stable (i.e., similar software units are developed or maintained over years), the major factor determining *defect content* (i.e., the number of defects inside the artifact when the QA activity starts) is the size of the artifact [23]. In such a context, we get a stable *defect density* (DD) value over our projects:

**Defect Density = Defect Content / Size   (1)**

In most environments, however, more factors exist besides artifact size that influence the defect content (e.g., developer experience, novelty of application, time pressure). The DC model captures the accumulated influence of these factors as the *defect density increase factor* (DDIF). Thus, defect density can be split into a *base defect density* ($DD_{base}$), which represents the hypothetical minimum defect density reachable in the considered context (best case), and a *defect density increase factor* describing the relative increase of this base defect density caused by the influence of the factors increasing the defect density (DDIF):

**Defect Content = Size · $DD_{base}$ · (1+DDIF)   (2)**

For illustrating equation (2), imagine artifacts checked in project A and B have the same size, but project A has a DDIF value of 0.0 (best case) and project B has a DDIF value of 0.3: then the artifact checked in project B has a 30% higher defect content ($DC_{Project2}$ = $DC_{Project1}$ · 1.3).

*Effectiveness Model:* The defect removal *effectiveness* of a QA activity can be defined as the number of defects found by the QA activity divided by the total number of defects in the product before the QA activity starts [22]. Like the defect density, the effectiveness of a QA activity depends on different influencing factors (e.g., experience of the QA team, understandability of the checked artifact, available tools). The accumulated impact of these factors on the effectiveness of a QA activity in a concrete project is captured by the *effectiveness improvement factor* (EIF). Therefore, like the actual defect density, the actual effectiveness can be split into two components: The *base effectiveness* ($Eff_{base}$) represents the hypothetical minimum effectiveness of the QA activity in the considered context (worst case), and the *effectiveness improvement factor* (EIF) describes the relative increase of this base effectiveness caused by the influence of effectiveness improvement factors.

**Effectiveness = $Eff_{base}$ · (1+EIF)   (3)**

For illustrating equation (3), imagine project A has an EIF value of 0.0 and project B has an EIF value of 0.2: then the QA activity in project B is 20% more effective than the activity in project A ($Eff_{Project2}$ = $Eff_{Project1}$ · 1.2).

### 3.1.2. Quantitative Causal Model

In general, providing estimates of the defect density increase factor (DDIF) and the effectiveness improvement factor (EIF) is a difficult task. We propose determining the defect density increase factor and the effectiveness improvement factor by building and applying a *quantitative causal model* for defect content and effectiveness.

Such a model consists of three components. First, it contains the most important influencing factors on *defect density* and *effectiveness* and their causal relationship in the considered context (example: Figure 2). Second, for each influencing factor, it provides a scale with usually four levels, which allows experts to characterize a project with respect to the factor (example: Table 3). Finally, for each factor, it captures the experts' judgment about the factor's impact on the number of detected defects in a quantitative way. The process we propose for building such a model (see Section 3.2) is similar to the process successfully applied in the cost estimation method CoBRA® [12] for building quantitative causal models for productivity.

### 3.1.3. Determine DDIF and EIF through Simulation

Since DDIF and EIF are determined based on expert judgment, which contains uncertainty, modeling DDIF and EIF as an exact value would be inappropriate. A single value cannot capture the uncertainty inherent in the usage of expert opinion. Therefore, DDIF and EIF are calculated and presented as probability distributions. For a project characterized by the level of each factor, Monte Carlo simulation [24] is applied to determine the DDIF and EIF probability distributions with the help of the quantitative causal model. Due to space restrictions, the approach for building a probability distribution from a given quantitative causal model and project characterization is not described in detail but can be found in [12] and [25].

## 3.2 Process: Quantitative Causal Model Building

The quantitative causal model captures the expert-opinion-based part of the HDCE model. This section gives a short overview of the process proposed for building such a quantitative causal model for defect content and effectiveness in a given context. The process consists of three major steps: (1) Define the application context and identify the most important influencing factor, (2) build a causal model based on these factors and define a rating scale for each factor in order to characterize projects with respect to the factor, and, finally, (3) quantify the impact of each factor on defect content or QA effectiveness based on expert judgment.

### 3.2.1. Identifying Most Important Influencing Factors

When considering defect density, usually some factors besides size exist that influence the number of defects in a product (e.g., developer experience, novelty of application, time pressure). The same holds for QA effectiveness. Factors of the first type will affect DDIF; factors of the second type will affect EIF. The relevant factors depend on the context in which the model should be applied and which should be defined before. Based on the context definition, the domain experts collect a list of relevant factors in a brainstorming session. *Relevant* means (1) the level of the factor varies across projects, (2) the level of the factor can be determined or at least reasonably judged for each project, and (3) the experts assume that the variation has a noticeable impact on defect content or QA effectiveness. The factor identification process should be supported by a list of factors that may be relevant in the context. Such a list can be extracted from the literature (e.g., [7]) or reused from an earlier application in a similar context. It should be used to check completeness (i.e., no relevant factor is missed).

After elicitation of all relevant factors, the factors are ranked by the experts regarding their importance. A factor is considered more *important* than another if its variation is assumed to be responsible for a higher variation in the number of defects detected by QA than the factor compared.

Usually, the most convenient way to get this ranking is a questionnaire where factors are separated into groups of 6 to 12 factors and each expert is asked to provide a ranking from 1 to n with respect to factor importance for the factors in each group. Groups with more than 12 factors are very difficult to rank for experts, since too many comparisons have to be performed. The rankings provided by the experts can be analyzed with the help of descriptive statistics (e.g., mean, min, max, standard derivation) and Kendall's coefficient of concordance [26] in order to identify the most important factors and determine the agreement between the experts regarding the importance of certain factors.

### 3.2.2. Building Causal Model and Defining Scales

In general, two options are possible for building the causal model based on the analysis results. The first one, the *rigorous-ranking-based* option, is to build the model considering only the analysis results, i.e., selecting the most important factor in each category and all factors whose mean rank is at most 10% higher than the mean rank of the most important factor. In this case, no interactions between factors are considered. The second, *discussion-based* option is to review the analysis results in a workshop with the experts in order to decide which factors should be included in the model. The second option provides the advantage that disagreements between experts with respect to the meaning or importance of factors can be articulated. However, this option typically requires significant logistical effort for the meeting and results in more complex causal models.

In any case, considering the experience with causal models in cost estimation [14], 4 to 6 defect content and 4 to 6 effectiveness factors are a reasonable number of factors for a DCE causal model.

In order to characterize the value of an influencing factor for a project, scales with levels from 0 to 3 are used. The description for level 0 should describe a realistic situation in the considered context where the factor leads to a minimal increase in the number of defects found; the description for level 3, on the other hand, should describe a realistic situation in the context where the factor results in a maximal increase in the number of defects found. *Note:* Based on this definition, level 3 is the *worst case* for defect content factors (high defect content) and the *best case* for effectiveness factors (high QA effectiveness). The descriptions for level 1 and 2 should be defined in such a way that the impact of the factor on the number of defects found increases (if possible) in a linear manner from 0 via 1 and 2 to level 3. Defining a context-specific and realistic best case and worst case is essential because in the next step, experts have to determine the quantitative impact of each factor by comparing the best case with the worst case.

### 3.2.3. Determining Impact of Influencing Factors

The influence of a factor is defined as a relative percentage increase of defect content or QA effectiveness (i.e., the number of defects found by the QA activity). For each defect content factor, experts are asked to provide an estimate for the expected increase in defect content when the considered factor has the worst possible value (level 3) compared to the best case (level 0). For each effectiveness factor, experts are asked to provide an estimate for the relative number of more defects found when the considered factor has the best possible value (level 3) compared to the worst case (level 0). Considering experts' uncertainty and the uncertainty of a factor's concrete impact in the context, they are asked for a minimum, most likely, and maximum estimate (called *multiplier*). Asking for these three values is a common technique in quantitative risk analysis [27].

### 3.3 HDCE Model Application

#### 3.3.1. QA Planning: Evaluate Quality Risk

The quantified causal model can be used during the QA planning phase to evaluate the remaining quality risk based on the planned QA activities. The basic idea is to benchmark the current project against historical projects with respect to expected defect density and QA effectiveness.

The required information includes expert-based characterizations of the current and at least 4 to 5 historical projects with respect to the level of influencing factors defined in the quantitative causal model. The characterization of the projects allows calculating the DDIF and EIF values for each project by MC simulation. In the following, we assume that effort spent on the QA activity is adapted to the size of the checked artifact (e.g., for each document page, two person-hours are available for the planned QA activity).

The result is visualized in a Cartesian coordinate system where the zero point of the x-axis (named relative defect density) is equal to the mean DDIF over all historical projects ($DDIF_{avg}$) and the zero point of the y-axis (named relative effectiveness) is equal to the mean EIF over all historical projects ($EIF_{avg}$). An example can be found in Figure 3. Based on the relative defect density and QA effectiveness of the actual project when compared to the historical projects, the actual project can be found in one of the four quadrants (Q1) to (Q4) of the coordinate system:

*(Q1) High defect density and effectiveness* usually means no major quality risk, since the relative number of defects that slipped through the QA activity is low because of the high effectiveness.

*(Q2) Low defect density but high effectiveness* usually means a very low quality risk, but can also mean an inappropriately high QA intensity with respect to the defect density expected (resulting in unnecessary costs).

*(Q3) Low defect density and effectiveness* might also not mean major quality risks, since the relative number of defects that slipped through the QA activity is low (because of the low defect density).

*(Q4) High defect content but low effectiveness* means a quality risk, since a relatively high number of defects can slip through the QA activity and result in potentially low product quality.

#### 3.3.2. QA Controlling: Estimating Defects Found

Besides QA planning, the quantified causal model can be used to predict the expected number of defects found by QA activities. The value can be used to control the QA process, i.e., if significantly fewer or more defects are found, the reason must be identified. The reason may be a derivation from the expected defect density or a derivation from the planned effectiveness of the QA activity.

In the case of higher effectiveness or lower defect density, this is usually not a problem, but in the case of lower effectiveness or higher defect density, this may be a quality risk leading to low final product quality or high correction costs in later project phases. Therefore, countermeasures should be initiated to handle the identified higher risk.

Since for each historical project j, $DDIF_j$ and $EIF_j$ can be calculated based on the characterization of the project with respect to the factors in the quantitative causal model, we can use these values together with the artifact size and number of defects found to calculate the $(DD_{base} \cdot Eff_{base})_j$ value using Equation 5:

$$(DD_{base} \cdot Eff_{base})_j = DF_j / (Size_j \cdot (1+ DDIF_j) \cdot (1+EIF_j)) \quad (5)$$

Based on our model assumptions, this value should have low variation across projects in the specified context because variations in defect density and effectiveness are encapsulated in the DDIF and EIF values. Therefore, the median of the *n* historical $(DD_{base} \cdot Eff_{base})_{j=1..n}$ values is used to provide an estimate for the $(DD_{base} \cdot Eff_{base})$ value of the actual project. The median and not the mean is used because the median is less affected by outliers than the mean.

If a sufficient number of historical projects is available for model building – around 10 projects – (robust) regression analysis can also be an alternative for determining the $(DD_{base} \cdot Eff_{base})_{est}$ value for the actual project.

When $(DD_{base} \cdot Eff_{base})_{est}$ is determined, the expected number of defects found (DF) can be calculated by applying the DCE equation for the actual project:

$$DF = Size \cdot (1+DDIF) \cdot (1+EIF) \cdot (DD_{base} \cdot Eff_{base})_{est} \quad (6)$$

### 4. Case Study

#### 4.1. Context of the Study

The presented case study was performed in the context of independent verification and validation (IV&V) of mission-critical, on-board space system software. The objective was to build a hybrid prediction model for the IV&V activities performed during the software requirements analysis phase (SRA). These SRA activities are mainly performed by document review. Depending on the project's situation, model checking is sometimes performed.

The data applied for constructing and evaluating the HDCE model came from 5 historical projects where software IV&V was performed during SRA. These projects were international collaborative projects (e.g., JEM and the HII-A Transfer Vehicle) and other spacecraft projects requiring an expense budget for IV&V[28].

Table 1 shows information about the domain experts who participated in building the model, by answering the prepared questionnaires, among other things.

**Table 1. Involved domain experts**

| Expert ID | Role of the expert | Experience [#years] | |
|---|---|---|---|
| | | Domain | IV&V |
| 1 | SPA/IV&V | 20 | 2 |
| 2 | SPA/IV&V | 25 | 2 |
| 3 | SW Safety/IV&V | 17 | 10 |
| 4 | IV&V | 17 | 5 |
| 5 | IV&V | 13 | 9 |
| 6 | IV&V | 5 | 5 |
| 7 | SW Safety/SPA/IV&V | 8 | 8 |

SPA: Software Product Assurance, SW Safety: Software Safety

## 4.2. Overall Study Goals and Research Questions

The primary goal of the study is to evaluate the applicability of the HDCE method and the usefulness of the resulting model by answering the following research questions:

**RQ1:** Is a method that combines expert and measurement data also applicable in the context of defect content and effectiveness prediction?

**RQ2:** How much expert involvement (in terms of effort) is required to build the quantitative DCE causal model?

**RQ3:** Does the model built provide plausible results for quality risk assessment when planning QA activities?

**RQ4:** How useful is the model for predicting the number of defects found by a QA activity?

Research questions RQ1 and RQ2 are answered in Section 4.3 by applying the proposed HDCE method to build a quantitative causal model for defect content and QA effectiveness. Afterwards, the resulting model is evaluated in Section 4.4 regarding its usefulness for QA planning and controlling (RQ3 and RQ4) using historical project data.

## 4.3 Evaluation: Model Building (RQ1&RQ2)

### 4.3.1. Study Objective and Design

The ability to build a context-specific causal model for defect content and effectiveness primarily depends on two prerequisites: (1) experts able to identify a set of factors they consider to have the highest impact in their context, and (2) a certain degree of agreement between the judgments of the domain experts.

Whether an identified factor really has the assumed impact in the considered context cannot be determined directly in the case study, but can only be measured indirectly by checking the improvement of estimation accuracy when applying the HDCE model containing the identified factors. The estimation accuracy is evaluated in detail in Section 4.4 (RQ4) with the help of legacy data.

### 4.3.2. Study Instrumentation and Execution

*Identification of relevant factors:* In order to identify the influencing factors, we used results from historical discussions in the IV&V community [29] as input. In addition, we included domain-related factors for defect content and effectiveness from the literature [7]. Initially, we had a set of more than 100 identified factors. From this set, we carefully selected context-relevant factors by discussing the factors with a domain expert to decide whether (1) the level of the factor varies in the context, (2) the variation of the factor level has an impact on defect content or IV&V effectiveness, and (3) the level of the factor can be determined by IV&V personnel. Finally, we identified 21 relevant defect content and 18 relevant effectiveness factors (see Table 2).

*Identification of most important factors:* In a first questionnaire, seven domain experts were asked to rank the factors with respect to their importance (see Section 3.2.1). In order to make the ranking easier for the experts, the factors were not only separated by the categories defect content and effectiveness, but also by the categories 'product', 'project', and 'process & personnel' (see Table 2). This allowed the experts to rank the factors in each of the six categories individually. The mean of the expert-based rankings that a factor received is presented in the table, where lower rankings mean higher importance. The agreement between the experts is measured by *Kendall's coefficient of concordance* (W) [26]. W ranges between 0 and 1, with 1 corresponding to perfect agreement between the experts and 0 indicating no agreement or independence of the sample.

*Factor selection for the causal model:* Since no model building workshop with the domain experts could be performed, we decided to select factors based solely on statistics (see *rigorous-ranking-based* option described in Section 3.2.2). In Table 2, the identified factors are highlighted.

**Table 2. Identified relevant influencing factors**

| Category : Defect Content - Product | | W=0.3778* |
|---|---|---|
| Imp. | Factor Name | Mean** |
| 1 | Novelty to developer | 4.143 |
| 2 | Number and complexity of external interface | 4.143 |
| 3 | Autonomous | 4.286 |
| 4 | Required failure tolerance | 5.429 |
| 5 | Requirement's assumption | 5.429 |
| 6 | Number of component decompositions | 6.142 |
| 7 | Time criticality | 6.286 |
| 8 | Hardware architecture | 6.571 |
| 9 | Role of functionality | 6.714 |
| 10 | Sub-architecture | 7.571 |
| 11 | Legacy part | 9.714 |
| 12 | Memory size | 11.429 |
| **Category : Defect Content - Project** | | **W=0.3143** |
| 1 | Number of stakeholder/user organizations | 1.429 |
| 2 | Involvement of customer in development | 2.714 |
| 3 | Developer's stress | 2.857 |
| 4 | Size of developer's project team | 3.000 |
| **Category : Defect Content - Process & Personnel** | | **W=0.4531*** |
| 1 | Project complexity | 1.429 |
| 2 | Schedule adherence | 2.174 |
| 3 | Management of developer | 3.000 |
| 4 | Developer's requirements analysis | 3.571 |
| 5 | Developer's knowledge about tools | 4.286 |
| **Category : QA Effectiveness - Product** | | **W=0.1230** |
| 1 | Consistent terminology of documentation | 2.714 |
| 2 | Documentation completeness | 2.857 |
| 3 | Type of language | 3.386 |
| 4 | Documentation of exceptional behavior | 3.714 |
| 5 | Documentation structure | 4.143 |
| 6 | Figures/charts in documentation | 4.286 |
| **Category : QA Effectiveness - Project** | | **W=0.2257** |
| 1 | Relationship between developer and IV&V | 2.714 |
| 2 | Management of document change | 2.714 |
| 3 | Involvement of customer in development | 3.429 |
| 4 | Disclosure of electronic file | 3.429 |
| 5 | Experience of IV&V manager | 3.571 |
| 6 | Transparency to stakeholder | 5.143 |
| **Category : Effectiveness - Process & Personnel** | | **W=0.4752*** |
| 1 | IV&V's knowledge of system/operation | 1.429 |
| 2 | Support from developer (supplier) | 2.517 |
| 3 | Supplier performs FTA | 4.429 |
| 4 | IV&V team relationship | 4.714 |
| 5 | Size of the IV&V team for review after RA | 5.000 |
| 6 | Size of the IV&V team for risk analysis (RA) | 5.714 |
| 7 | Supplier performs FMEA | 5.857 |
| 8 | IV&V's experience with tool | 6.286 |

\* significant at α=0.05   \*\* of the rankings provided by the experts

*Factor scale definition:* Based on the discussion between a domain expert and a measurement expert, the final scales for the factors were defined. For each factor, the scales provide a context-specific wording to define the factor's levels 0 to 3 (see Table 3). For level 0 and level 3, a concrete worst and best case situation was chosen based on the experience from historical projects in the context.

**Table 3. Example of context-specific factor scale**

| Factor name: Number of stakeholder/user organizations | | |
|---|---|---|
| Level | Description | Selection |
| 0 | Stakeholders are customer and supplier. | |
| 1 | … customer, supplier, and user. | X |
| 2 | … customer, supplier, and international partner. | |
| 3 | … customer, supplier, several users, and international partner. | |

*Factor quantification with questionnaire:* A second questionnaire supported the seven experts quantifying the impact of the factors on defect content and IV&V effectiveness by providing so-called multipliers for each factor (see Section 3.2.3). The answers of one expert had to be removed because of invalid answers caused by a misunderstanding of the questionnaire.

### 4.3.3. Results: Applicability of the method (RQ1)

Table 2 includes Kendall's coefficient of concordance $W$ for each category. We got significant agreement for half of the categories. We observed the highest disagreement between experts for product-related factors influencing effectiveness ($W=0.123$) and the highest agreement for process- and personnel-related factors influencing effectiveness ($W=0.475$). Finally, we were able to build an initial causal model based on the defined factor selection criterion (following the rigorous-ranking-based approach). The resulting IV&V DCE causal model includes five defect content factors and five IV&V effectiveness factors (Figure 2).

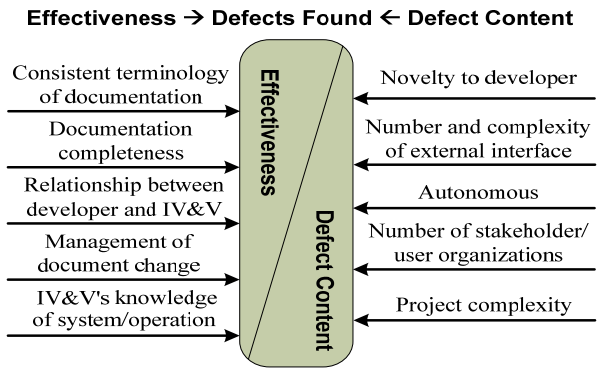

**Figure 2. Resulting DCE causal model**

### 4.3.4. Results: Effort for model building (RQ2)

The identification of relevant factors was based on a discussion of an existing set of factors between one domain expert and one measurement expert. Relevant factors were selected, categorized, and the best and worst cases were quoted. Thus, additional experts were only involved in answering the two questionnaires for factor ranking and quantification. The median effort for the ranking questionnaire was one hour, and varied between 30 and 90 minutes. To answer the quantification questionnaire, the median effort was 40 minutes and varied between 30 and 60 minutes. On average, the total effort per expert was 112 minutes; this includes time required for requests.

### 4.4 Evaluation: Model Application (RQ3 & RQ4)

After the quantitative causal model was built in the case study, available legacy data (and, to some extent, expert opinion) were used to evaluate the usefulness of the model for planning and controlling QA activities.

#### 4.4.1. Study Objectives and Design

Since for the historical projects, no data was available about the number of defects that slipped through the QA activities in SRA, the usefulness of the model with respect to *planning QA* (RQ3) by providing quality risk analysis could not be evaluated directly. Therefore, the plausibility of the model results was evaluated by checking the following two hypotheses:

**H3.1:** Higher effectiveness results in more defects found per size unit, i.e., if a project has a higher relative effectiveness (EIF) compared to a second project and the second project has no higher relative defect density (DDIF), then the *number of defects found per artifact size unit (DF/Size)* in the first project is higher than in the second project. The same holds for defect density; higher defect density results in more defects found per size unit.

**H3.2:** The experts agree on the quality risk assessment results provided by the model for the historical projects.

The usefulness of the model for *controlling QA* (RQ4) by providing an indicator or thresholds for the number of defects expected to be found by the QA activity is checked by determining the prediction error of the HDCE model, comparing the prediction error with the prediction error of applicable data-based prediction models, and checking the usefulness of each component of the model (defect content factors, effectiveness factors, size information).

**H4.1:** The *estimation accuracy* of the DCE model is higher than the accuracy of a model using only available measurement data (i.e., not including expert judgment).

**H4.2:** Each component of the estimation model (i.e., artifact size, EIF, and DDIF) contributes to the *estimation accuracy* of the HDCE model.

#### 4.4.2. Study Instrumentation and Execution

To apply the model for planning and controlling QA and check our hypotheses, the used constructs had to be operationalized, data had to be collected, and evaluation procedures needed to be defined (e.g., by selecting appropriate statistical tests):

*Number of defects found (DF)* is measured as the number of issues that had been found by the IV&V supplier in SRA. The total number of issues is counted, even if certain issues are unrelated to problems of the development project.

*Size of artifact (Size)* is measured as the number of pages of the document checked in SRA by the IV&V supplier.

*Project characterization* with respect to the influencing factors is provided on the four-level scales defined by the causal model for each factor (Section 3.2.2).

*Estimation accuracy* refers to the nearness of an estimate to the true value. In order to remain comparable to other estimation studies, we used common estimation error measures and accuracy measures [30], such as *relative error* (*RE*) and *mean magnitude of relative error* (*MMRE*). In the following, we consider a model to be more accurate than another one if its MMRE value is lower.

**Data Collection:** The number of *defects found* and the *artifact size* were collected from IV&V activities in SRA for five available historical projects (A to E) by checking the issue reports and SRA documents. The *project characterization* was conducted by one expert who knew the five projects. The result was later discussed with a second expert in order to check the validity of the characterization.

**Procedure for QA Planning (RQ3):** Based on the project characterization provided by the experts, the resulting distributions for DDIF and EIF were determined by Monte Carlo simulation for each historical project using the freely available CoBRIX tool [31]. Using the DDIF and EIF distributions a quality risk chart was created and experts were asked to provide their retrospective option about the relative quality risk in the considered historical projects.

**Procedure for QA Controlling (RQ4):** To evaluate the estimation accuracy, we used *leave-one-out cross-validation* [32], justified by the low number of projects available for prediction model building and validation. In our case, this meant that four of the five projects were used to build a prediction model that was applied to estimate the defects found in the fifth 'unknown' project. This procedure was repeated for each of the five projects and the resulting MMRE was calculated. Since Kitchenham [33] criticized MMRE as a measure of model accuracy, but MMRE is the de-facto standard, we decided to present box plots of RE values for visual examination of the results as recommended by [33] in addition to MMRE.

The model was compared to the two reasonable prediction models (H4.1) that can be built with the available project data, namely defects found and artifact size. (1) The $DF_{only}$ model assumes an equal number of defects found for each project and takes the median of the defects found in the historical projects to predict the number for the actual one. (2) The *DF+Size* model assumes stable defect density and effectiveness for all projects. Therefore, it uses the median defect density of the historical projects to estimate the defect density of the actual project and then multiplies the estimated defect density with the artifact size of the actual project to predict the number of defects found.

To evaluate the contribution of the three components of the model (H4.2) – defect content factors (DDIF), effectiveness factors (EIF), and artifact size – we evaluated the accuracy of the model when a certain component was not considered. This was done by setting DDIF=0 (*w/o DDIF*), EIF=0 (*w/o EIF*), or size=1 (*w/o size*) for all projects, respectively.

The significance of the observed differences in the magnitude of relative error (MRE) was checked by a two-sided *Wilcoxon test* (i.e., the non-parametric pendant of the paired t-test, which was not applicable since we could not assume a normal distribution). As significance level, we chose .05.

### 4.4.3. Result & Interpretation: QA Planning (RQ3)

The quality risk analysis resulting from the model application is presented in Figure 3. The five dots represent the five historical projects A to E. The DDIF and EIF values of the project, which represent the mean values of the respective distribution, are scaled by a fixed factor (f) for confidentiality issues. This means, for example, for project A:

*relative defect density* = $(DDIF_A - DDIF_{average}) \cdot f$
*relative effectiveness* = $(EIF_A - EIF_{average}) \cdot f$

Following hypothesis H3.1, one would assume when looking at Figure 3 that the number of *defects found per page* (*DF/P*) increases for the three projects located on nearly the same effectiveness level with increasing defect density. This is confirmed by available DF and size data: $DF/P_E > DF/P_B > DF/P_D$. Project A with high relative effectiveness compared to the remaining projects and project C with high relative defect density should both have a significantly higher DF/P value compared to E, B, and D. This is also confirmed by available DF and size data: $DF/P_A$ and $DF/P_C \gg DF/P_D$. This supports the plausibility of the model results based on available historical data (H3.1). The absolute DF/P values for the projects are not presented here because of confidentiality issues.

In addition, the experts were asked for their retrospective opinion about the risk of the five projects and agreed on the results that project C was the most risky one and the risk in project A (although many defects were found) was not assessed to be high, especially because of effective IV&V. Project B and E were considered in the routine catch of IV&V experience in the organization. These results support the plausibility of the model results based on retrospective expert opinion (H3.2).

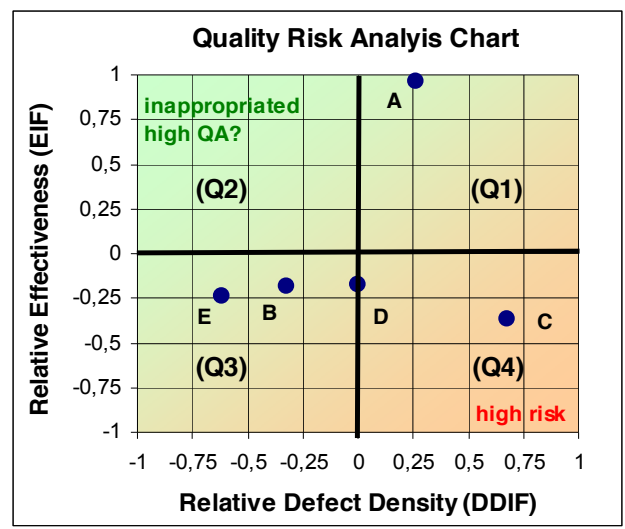

**Figure 3. Historical projects in risk chart**

### 4.4.4. Result & Interpretation: QA Controlling (RQ4)

The accuracy achieved by the HDCE model is measured by an MMRE of 29.6%. This value seems good for an initial model when compared to the results of initial cost estimation models built by applying CoBRA [14] (initial model 107%, first iteration 32%), but improvable (fourth iteration 14%). We performed no iterations to improve the model's accuracy due to time constraints and the fear of overfitting the model with respect to the limited number of historical projects.

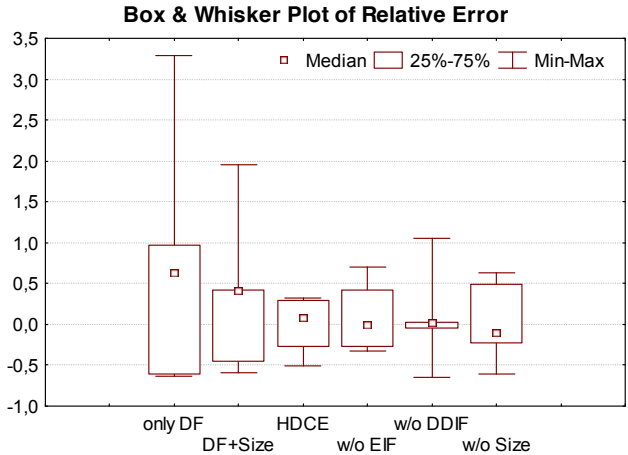

**Figure 4. Relative error of prediction models**

Independent of these considerations, H4.1 can be confirmed, since both data-based models ($DF_{only}$ and $DF+Size$) provide estimates with significantly higher estimation errors ($MMRE_{DFonly}$ = 122.8%, $MMRE_{DF+Size}$ = 76.5%), with significance being tested using the Wilcoxon test at a significance level of .05. Reduced estimation error can also be observed visually by checking the box plots in Figure 4.

In the same way, H4.2 can be confirmed, since a model not considering the factors influencing effectiveness would result in reduced accuracy ($MMRE_{w/oEIF}$ = 34.2%), as would a model not considering factors influencing defect content ($MMRE_{w/oDDIF}$ = 35.4%) or artifact size, i.e., using only expert opinion ($MMRE_{w/oSize}$ = 41.2%). These results show that not only the expert-based component (H4.1), but also the measurement component contributes to model accuracy.

## 5. Threat to Validity

In this section, we discuss possible threats to the validity of this study. We referred to [15] for categorizing the threats:

**Low statistical power** (conclusion validity): Only a small number of historical projects (5) was available for building and evaluating the HDCE model.

**Reliability of measure** (conclusion validity): Individual experts' criteria influence the number of issues reported that were used as a measure for defects found. In addition, we did not perform an introductory workshop for the questionnaires; therefore, there might be some bias of understanding regarding the purpose of the questionnaires.

**Mortality of Q&A** (internal validity): During the second round of questionnaires, answers from one expert had to be removed because of invalid answers caused by a misunderstanding of the questionnaire.

**Homogeneity of experts** (external validity): In this study, we built the causal model based only on the answers provided by IV&V personnel from one organization. If we consider another organization, the result would be different, since the built model is context-specific.

## 6. Summary and Conclusions

In this paper, we pointed out the lack of empirical validation of context-specific prediction methods for planning and controlling quality assurance activities during early phases of the software lifecycle when only limited historical data points and measurement data are available. We also pointed out that the defect content of the investigated artifact and the effectiveness of quality assurance activities are context-specific and affected by several influencing factors.

Therefore, we proposed a hybrid defect content and effectiveness (HDCE) prediction method that allows building context-specific models that consider the most relevant influencing factors in the context. The HDCE method combines available historical project data and expert judgment encapsulated in a reusable quantitative causal model for factors influencing defect content and effectiveness.

The *applicability of the method* (RQ1) was shown by conducting a case study in the context of IV&V for the software requirements analysis phase of critical software in the space development domain. To build the required causal model, the knowledge of seven domain experts was elicited basically with the help of questionnaires, which required, on average, 112 minutes per expert in total (RQ2).

The usefulness of the resulting causal model was evaluated using measurement data collected for five historical projects (artifact size, number of defects found). The evaluation results suggest the usefulness of the model for *QA planning* (RQ3) by identifying projects with high quality risk. Moreover, this also holds for *QA controlling* by providing statistically significant better estimates for the number of defects expected to be found than models using only measurement data (RQ4).

In addition, from the perspective of the IV&V experts, the mapping of the probability distribution for the five historical projects showed the predicted risk of IV&V strategy and confirmed the experts' impression regarding historical evaluated projects. Consequently, the proposed hybrid prediction model will be integrated into the IV&V planning and monitoring procedures of JAMSS to support QA activities during the software requirements phase.

As a future direction, the HDCE prediction model should be expanded by collecting and including defect data from testing phases to determine defect slippage during earlier phases. This expansion would serve to make the model more precise and would allow predicting absolute defect content and effectiveness values. Furthermore, we plan to expand our model to predict risk exposure during the operational phase (after release), including severity of defects.

## Acknowledgement

We would like to thank the development project staff and IV&V staff from the JAXA Engineering Digital Innovation Center (JEDI) at the Japanese Aerospace Exploration Agency (JAXA), where we conducted the case study to construct the hybrid prediction model. We would like to thank the staff of JAMSS, who greatly contributed by answering the questionnaires and giving us historical experience data. Finally, we would like to thank Adam Trendowicz and Marcus Ciolkowski from Fraunhofer IESE for the initial review of the paper and Sonnhild Namingha for proofreading. Parts of this work have been funded by the BMBF SE2006 project TestBalance (grant 01 IS F08 D).